\documentstyle[preprint,revtex]{aps}

\begin{document}

\begin{title} Gauging the octonion algebra \end{title}
\author{A.K. Waldron and G.C. Joshi}
\begin{instit} Research Centre for High Energy Physics, \\ University of
Melbourne, Parkville, Victoria 3052, Australia \end{instit}

\begin{abstract}
By considering representation theory for non-associative algebras we construct
the fundamental and adjoint representations of the octonion algebra. We then
show how these representations by associative matrices allow a consistent
octonionic gauge theory to be realized. We find that non-associativity implies
the existence of new terms in the transformation laws of fields and the kinetic
term of an octonionic Lagrangian.

\end{abstract}
\section{Introduction}

The aim of this work is to genuinely gauge the octonion algebra as opposed to
relating properties of this algebra back to the well known theory of Lie Groups
and fibre bundles. Typically most attempts to utilise the octonion symmetry in
physics have revolved around considerations of the automorphism group $G_{2}$
of the octonions and Jordan matrix representations of the
octonions~\cite{quarky people}. Our approach is more simple since we provide a
spinorial approach to the octonion symmetry. Previous to this work there were
already several indications that this should be possible. To begin with the
statement of the gauge principle itself ``{\em no theory shall depend on the
labelling of the internal symmetry space coordinates}'' seems to be independent
of the exact nature of the gauge algebra and so should apply equally to
non-associative algebras.

 The octonion algebra is an alternative algebra (the associator $\{
x^{-1},y,x\} =0$ always) so that the transformation law for a gauge field
$\Gamma_{\mu}\rightarrow\Gamma '_{\mu}=
U\Gamma_{\mu}U^{-1}-\frac{i}{g}(\partial_{\mu}U)U^{-1}$ is well defined for
octonionic transformations $U$.

In analogy with the isomorphisms: unit quaternions $\leftrightarrow SU(2)$ and
unit complex numbers $\leftrightarrow U(1)$, the unit octonion algebra is of
premium interest. Topologically the unit octonions are an $S^{7}$ which is a
parallizable manifold. This has allowed Nahm~\cite{Nahm}  to show that an
octonionic Yang Mills theory does indeed exist. Further Lukierski and
Minnaert~\cite{Lukierski} have shown that although the unit octonions do not
associate they can still be treated as a soft Lie
Algebra~\cite{Sohnius}~\cite{fn1} with structure constants varying as pure
torsion on $S^{7}$.

The unique place of the octonion algebra as the last real division algebra in
the Hurwitz theorem sequence: reals, complex, quaternions and octonions, is
tantalising in view of the myriad of applications in physics of the former
three algebras. The current standard model group of particle physics follows
this sequence closely. Observe that given we could include the unit real
numbers $\{1,-1\} \cong Z_{2}$ to cover discrete symmetries only the $SU(3)$
group does not follow the unit Hurwitz real division algebra sequence. Since we
presently have nothing other than anthropic arguments for the actual Lie Group
chosen by nature it would be elegant if the Hurwitz theorem provided the
algebras of the standard model. It is known that the octonion algebra can be
used to generate all Lie Groups~\cite{GCJ} and $SU(3)$ is a subgroup of the
inner automorphism group of the octonions~\cite{quarky people} so an octonionic
standard model does not seem unreasonable {\em if} a ``genuine''
non-associative gauge theory could be
 developed.
\section{Bimodular representation theory}

Non-associative algebras and physics up till now have had a strained
relationship. Obviously non-associative numbers are difficult to manipulate.
However there are deeper inherent problems for non-associative symmetries in
particle physics. Firstly non-associativity implies the failure of unitarity in
octonionic quantum mechanics~\cite{fn2}. Secondly the edifice of modern gauge
theories has a profound basis in the theory of connections on fibre bundles.
One would expect any generalisation of the gauge principle to respect bundle
theory. Yet associativity of the structure group (=gauge algebra) is a direct
consequence of the associativity of the composition of maps (specifically the
transition functions giving the bundle its warped product structure). This is
central and incontrovertible for the fibre bundle formalism. Therefore in this
work we consider representation theory of non-associative yet alternative
algebras by associative matrices. In this way we circumvent the failure of
unitarity and need not who
leheartedly fling away the fibre bundle. From a philosophical viewpoint this is
acceptable since in physics it is never the abstract group that is of interest
but rather its representation. So let us now consider representation theory of
alternative algebras.

An alternative algebra $A$ is an algebra in which the associator
\begin{equation}\{ x,y,z\} = (xy)z-x(yz)\end{equation} is in general
non-vanishing. However an alternativity condition is fulfilled
\begin{equation}\{ x,y,z\} +\{ x,z,y\} =0 \label{altcdn}.\end{equation}
Equivalently any subalgebra of $A$ generated by two elements and the identity
is associative.

Now observe that an element $x\in A$ alone is not a well defined object. For if
$a \neq b$ then the triple product $axb$ could be either $(ax)b$ or $a(xb)$.
Therefore a single element $x \in A$ actually entails two objects
$\stackrel{\leftarrow}{x}$ and
$\stackrel{\rightarrow}{x}$ ($x$-left and $x$-right say) where
 \begin{equation} a\stackrel{\leftarrow}{x}b=(ax)b \mbox{ and
}a\stackrel{\rightarrow}{x}b=a(xb) .\end{equation}

Therefore representation theory for an alternative algebra involves two sets of
matrices/linear transformations. In this way non-associativity can be recast as
a ``doubling'' of the algebra.

Hence define a bimodular representation~\cite{Sorgsepp} of an alternative
algebra as a pair of linear mappings $(L,R)$ of $A$ into $L(V)$ the set of all
linear transformations of some vector space $V$ into itself, satisfying the
following relations:
\begin{eqnarray} &\mbox{If }& \left\{ \begin{array}{l} L:x \rightarrow L(x) \\
R:y \rightarrow R(y) \mbox{ where } x,y \in A \end{array} \right. \\
&\mbox{then }& \left\{ \begin{array}{ll} L(xy) = L(x)L(y)+ [R(x),L(y)] & \\
                                         R(xy)=R(x)R(y) + [L(x),R(y)] & \\
                                         \left[ L(x),R(y)] = [R(x),L(y)]
\right.
 & \forall x,y\in A .\end{array} \right.  \label{repreln}\end{eqnarray}

Further, since any two elements generate an associative subalgebra
\begin{equation} [L(x),R(x)] \equiv 0. \end{equation}

Notice that these relations mimic those for Lie groups except for the
commutator terms coupling the left and right sides of the representation to one
another. These terms provide new physics peculiar to alternative algebras.

\subsection{The fundamental representation}

Let us now consider the specific problem of constructing a fundamental
bimodular representation of the octonion algebra. The octonion algebra is an 8
dimensional real division algebra. Indeed it is the last in the sequence of the
4 unique real division algebras (Hurwitz theorem) reals, complex, quaternions
and octonions. A general octonion is \begin{equation} x=x_{0}e_{0}+ x_{1}e_{1}
+ \ldots + x_{7}e_{7}=x_{0}e_{0}+ x_{i}e_{i}. \end{equation} (We will follow
the usual convention whereby greek indices are used to include the identity
element of the algebra so that $\alpha ,\beta , \gamma , \ldots =0,\ldots,7$
and latin indices denote only the hypercomplex units so that
$i,j,k,\ldots=1,\ldots,7$.)

The units satisfy the following algebra
\begin{eqnarray} && e_{0}^{2} =e_{0} \\ && e_{0}e_{i} = e_{i}e_{0}=e_{i} \\  &&
e_{i}e_{j}= -\delta_{ij}+\epsilon_{ijk}e_{k} \mbox{   where } i,j,k =1,\ldots
,7 .\end{eqnarray}
$\epsilon_{ijk}$ is totally antisymmetric and $\epsilon_{ijk}=1$ for  $(ijk)=
(123)$, $(145)$, $(176)$, $(246)$, $(257)$, $(347)$, $(365)$ (each cycle
represents a quaternion subalgebra). Its dual is the totally antisymmetric rank
4 tensor $\epsilon_{ijkl}=1$ for  $(ijkl) =(1247)$, $(1265)$, $(2345)$,
$(2376)$, $(3146)$, $(3157)$, $(4576)$. Notice that for an $SU(2)$ theory no
such object is of relevance.
The dual structure constants also determine the associator
 \begin{equation} \{ e_{i},e_{j},e_{k} \} \equiv
(e_{i}e_{j})e_{k}-e_{i}(e_{j}e_{k})=-2\epsilon_{ijkl}e_{l}. \end{equation}

The $\epsilon_{ijk}$ structure constants obey the following
relations~\cite{fn3}
\begin{eqnarray}
\epsilon_{abi}\epsilon_{cdi}&=&
\delta_{ac}\delta_{bd}-\delta_{ad}\delta_{bc}+\epsilon_{abcd} \nonumber \\
\epsilon_{aij}\epsilon_{bij}&=& 6\delta_{ab}\nonumber \\
\epsilon_{ijk}\epsilon_{ijk}&=& 42.
\end{eqnarray}
Finally define the involution (octonionic conjugation) \begin{equation}
\bar{x}=x_{0}e_{0}-x_{i}e_{i}, \end{equation} so that
 \begin{equation} |x|^{2}=x \bar{x}=\bar{x}x=x_{0}^{2} + x_{i}x_{i}
\end{equation} and a unique inverse is defined
  \begin{equation} x^{-1}=\frac{\bar{x}}{|x|} \ \ \ \  (x\neq 0).\end{equation}

The fundamental representation should be the lowest dimensional faithful
representation by hermitian (or orthogonal) matrices.  On dimensional grounds
there are 7 octonionic units and therefore 14 matrix generators (15 including
the identity which being associative needs no doubling). The set of all
$N\times N$ hermitian matrices is generated by $N^{2}-1$ generators so the
lowest dimension we should look at is $N=4$ (or $N=8$ in the orthogonal case).
However it remains unclear how to construct a fundamental representation.
Therefore let us start with a well known associative case first and generalise.

The isomorphism between the quaternions and $SU(2)$ is well known. Hence we
provide a method of constructing the Pauli matrices (generators of the
fundamental representation of $SU(2)$) from the quaternion algebra.

The quaternion algebra is generated by 3 units $e_{1},e_{2},e_{3}$ (which we
interchangeably denote $i,j,k$). Allowing the indices $i,j,k$ to run from 1 to
3 in the octonion algebra above and replacing $\epsilon_{ijk}$ by the 3
dimensional Levi-Civita alternating symbol $\epsilon_{ijk}$ we obtain the
quaternion algebra. The quaternions are an associative but non-commutative
division algebra. Therefore inverses always exist but quotients are not unique.
Rather we must distinguish between the left quotient $x$ of two quaternions $a$
and $b$ where $bx=a$ and the right quotient $\tilde{x}$ where $\tilde{x}b=a$.
 The left division table for the quaternions is

\begin{center}
\begin{equation}\mbox{\begin{tabular}{c|cccc} $\div$ & $1$&$i$&$j$&$k$\\ \hline
$1$ & $1$ & $i$ &  $j$ & $k$ \\ $i$ & $-i$ & $1$ & $-k$ & $j$ \\ $j$ & $-j$ &
$k$ & $1$ & $-i$ \\
 $k$ & $-k$ & $-j$ & $i$ & $1$   \end{tabular}}\end{equation}
\end{center}

Take the coefficients of $1,i,j,k$ as the entries for the $4\times 4$ matrix
representatives of $1,i,j,k$ respectively. Denoting $1$ and $i$ by their
$2\times 2$ matrix representatives $\left( \begin{array}{cc} 1 & 0 \\ 0 & 1
\end{array} \right)$ and $\left( \begin{array}{cc} 0 & 1 \\ -1 & 0 \end{array}
\right)$ we obtain the Pauli matrices~\cite{pauli}
\begin{eqnarray}
1 & \longrightarrow & \left( \begin{array}{cc} \  1 &\  0 \\\  0 &\  1
\end{array} \right) =\bf{1} \nonumber \\
i & \longrightarrow & \left( \begin{array}{cc}\  i &\  0 \\\  0 & -i
\end{array} \right) = i\sigma_{1} \nonumber \\
j & \longrightarrow & \left( \begin{array}{cc}\  0 &\  1 \\ -1 &\  0
\end{array} \right) = i\sigma_{2} \nonumber \\
k & \longrightarrow & \left( \begin{array}{cc}\  0 &\  i \\ \ i & \ 0
\end{array} \right) =i\sigma_{3}. \end{eqnarray}

The same procedure can be followed for the right division table but of course
gives no independent results. We now apply the above derivation of the
fundamental representation from algebra to the octonion algebra. However since
the octonions fail to associate both the left and right division tables are
required. These are tabulated in appendix~\ref{appendix}.

In terms of the Pauli matrices ($\sigma_{1},\sigma_{2},\sigma_{3},\bf{1}$) we
then have the following orthogonal traceless left and right matrix
representatives

\begin{eqnarray}
e_{1} \longrightarrow &L(e_{1})=\left( \begin{array}{cccc} -i\sigma_{2} &\
 0&\  0 &\  0 \\
\ 0 & -i\sigma_{2} &\  0 & \ 0 \\
0 & \   0 & -i\sigma_{2}  &\   0 \\
\   0 &\  0 &\   0 &i\sigma_{2}  \end{array} \right) , &
R(e_{1})=\left( \begin{array}{cccc} -i\sigma_{2} &\       0&\  0 &\  0 \\
\ 0 & i\sigma_{2} &\  0 & \ 0 \\
0 & \   0 & i\sigma_{2}  &\   0 \\
\   0 &\  0 &\   0 &-i\sigma_{2}  \end{array} \right)  \nonumber \\
e_{2} \longrightarrow &L(e_{2})=\left( \begin{array}{cccc} \ 0 & -\sigma_{3}&\
0 &\  0 \\
\ \sigma_{3}  & \ 0 &\  0 & \ 0 \\
0 & \   0 & \ 0  & -\bf{1} \\
\ 0 &\  0 &\  \bf{1} &\ 0  \end{array} \right) , &
 R(e_{2})=\left( \begin{array}{cccc} \ 0 & -\bf{1}&\  0 &\  0 \\
\  \bf{1}  & \ 0 &\  0 & \ 0 \\
0 & \   0 & \ 0  & \ \bf{1} \\
\ 0 &\  0 &  -\bf{1} &\ 0  \end{array} \right)\nonumber \\
e_{3} \longrightarrow &L(e_{3})=\left( \begin{array}{cccc} \ 0 & -\sigma_{1}&\
0 &\  0 \\
\ \sigma_{1}  & \ 0 &\  0 & \ 0 \\
0 & \   0 & \ 0  & -i\sigma_{2} \\
\ 0 &\  0 &  -i\sigma_{2}&\ 0  \end{array} \right), &
R(e_{3})=\left( \begin{array}{cccc} \ 0 & -i\sigma_{2}&\  0 &\  0 \\
 -i\sigma_{2}  & \ 0 &\  0 & \ 0 \\
0 & \   0 & \ 0  & i\sigma_{2} \\
\ 0 &\  0 &  i\sigma_{2}&\ 0  \end{array} \right)  \nonumber \\
e_{4} \longrightarrow &L(e_{4})=\left( \begin{array}{cccc} \ 0 & \
0&-\sigma_{3} &\  0 \\
\ 0 & \ 0 &\  0 & \ \bf{1} \\
\ \sigma_{3} & \   0 & \ 0  & \ 0 \\
\ 0 &-\bf{1} &\ 0&\ 0  \end{array} \right) , &
R(e_{4})=\left( \begin{array}{cccc} \ 0 & \ 0&-\bf{1} &\  0 \\
\ 0 & \ 0 &\  0 & -\bf{1} \\
\ \bf{1} & \   0 & \ 0  & \ 0 \\
\ 0 &\ \bf{1} &\ 0&\ 0  \end{array} \right)  \nonumber \\
e_{5} \longrightarrow &L(e_{5})=\left( \begin{array}{cccc} \ 0 & \
0&-\sigma_{1} &\  0 \\
\ 0 & \ 0 &\  0 & i\sigma_{2} \\
\ \sigma_{1} & \   0 & \ 0  & \ 0 \\
\ 0 &i\sigma_{2} &\ 0&\ 0  \end{array} \right),&
 R(e_{5})=\left( \begin{array}{cccc} \ 0 & \ 0&-i\sigma_{2} &\  0 \\
\ 0 & \ 0 &\  0 & -i\sigma_{2} \\
 -i\sigma_{2} & \   0 & \ 0  & \ 0 \\
\ 0 &-i\sigma_{2} &\ 0&\ 0  \end{array} \right)  \nonumber \\
e_{6} \longrightarrow &L(e_{6})=\left( \begin{array}{cccc} \ 0 & \ 0&\ 0
&-\bf{1} \\
\ 0 & \ 0 &-\sigma_{3} & \ 0 \\
\ 0 & \ \sigma_{3} & \ 0  & \ 0 \\
\ \bf{1} &\ 0 &\ 0&\ 0  \end{array} \right), &
R(e_{6})=\left( \begin{array}{cccc} \ 0 & \ 0&\ 0 &-\sigma_{3} \\
\ 0 & \ 0 &\ \sigma_{3} & \ 0 \\
\ 0 & -\sigma_{3} & \ 0  & \ 0 \\
\ \sigma_{3} &\ 0 &\ 0&\ 0  \end{array} \right)  \nonumber \\
e_{7} \longrightarrow &L(e_{7})=\left( \begin{array}{cccc} \ 0 & \ 0&\ 0
&-i\sigma_{2} \\
\ 0 & \ 0 &-\sigma_{1} & \ 0 \\
\ 0 & \ \sigma_{1} & \ 0  & \ 0 \\
-i\sigma_{2} &\ 0 &\ 0&\ 0  \end{array} \right),&
 R(e_{7})=\left( \begin{array}{cccc} \ 0 & \ 0&\ 0 &-\sigma_{1} \\
\ 0 & \ 0 &\ \sigma_{1} & \ 0 \\
\ 0 & -\sigma_{1} & \ 0  & \ 0 \\
\ \sigma_{1} &\ 0 &\ 0&\ 0  \end{array} \right)
\end{eqnarray}

Notice the elegant sequence: quaternions $\leftrightarrow$ represented by $2
\times 2$ complex matrices, octonions $\leftrightarrow$ represented by $4
\times 4$ quaternionic matrices.

\subsection{The regular representation}\label{adj}

Typically the regular (or adjoint) representation for Lie groups is derived
from the Jacobi identity \begin{equation}
[A,[B,C]]+[C,[A,B]]+[B,[C,A]]=0.\end{equation}
The structure constants of the algebra then become the matrix representatives.
However the Jacobi identity fails for non-associative $A,B,C$. Therefore we
must look for an alternative construct.

Define the Jacobi function \begin{equation} {\cal J}\! ac
(x,y,z)=[x,[y,z]]+[z,[x,y]]+[y,[z,x]] \end{equation} which vanishes always for
associative $x,y,z$.
It is easy to check that the alternativity condition~\ref{altcdn} gives
\begin{equation} {\cal J}\! ac(x,y,z)+{\cal J}\! ac(x,z,y)=0.
\label{jac}\end{equation}
Let the units of our alternative algebra $A$ be $\{ e_{i}\}$
($i=1,\ldots,\mbox{dim}(A)-1$). Define the structure constants
$\epsilon_{ij}^{k}$  of $A$ in the usual way
\begin{equation}[e_{i},e_{j}]=2\epsilon_{ij}^{k}e_{k}.\end{equation}
Substituting these into~\ref{jac}
\begin{eqnarray} 0&=&{\cal J}\! ac(e_{i},e_{j},e_{k})+{\cal J}\!
ac(e_{i},e_{k},e_{j})\\
   	           &=&
4[\epsilon_{il}^{m}\epsilon_{jk}^{l}+\epsilon_{kl}^{m}c_{ij}^{l}+\epsilon_{jl}^{m}\epsilon_{ki}^{l}+\epsilon_{jl}^{m}\epsilon_{ki}^{l}+\epsilon_{jl}^{m}\epsilon_{ik}^{l}+
\epsilon_{kl}^{m}\epsilon_{ji}^{l}]e_{m} \end{eqnarray}
Which holds for any $e_{m}$ so that
\begin{equation}\epsilon_{il}^{m}\epsilon_{jk}^{l}-\epsilon_{jl}^{m}\epsilon_{ik}^{l}+\epsilon_{il}^{m}\epsilon_{kj}^{l}-\epsilon_{lj}^{m}\epsilon_{ik}^{l}=2\epsilon_{ij}^{l}\epsilon_{lk}^{m}\end{equation}

Now replacing the structure constants with fixed $i$ and $j$ with matrices
suggestively labelled
\begin{equation}\left\{ \begin{array}{l} {\cal L}_{i}=[\epsilon_{i}]_{l}^{m}
\leftrightarrow \epsilon_{il}^{m} \\ {\cal
R}_{j}=[\tilde{\epsilon}_j]_{l}^{m}\leftrightarrow \epsilon_{lj}^{m}\end{array}
\right. \end{equation} we have \begin{equation} [{\cal L}_{i},{\cal
L}_{j}]+[{\cal L}_{i},{\cal R}_{j}]=2\epsilon_{ij}^{l}{\cal L}_{l} .
\label{adjrepL} \end{equation}

Notice that for an associative algebra the commutator $[{\cal L}_{i},{\cal
R}_{j}]=0$ and we then have the usual adjoint matrix representation for Lie
groups. Further equation ~\ref{adjrepL} is exactly the bimodular representation
relation~\ref{repreln} written in terms of commutators. The equivalent result
for the right matrices ${\cal R}_{i}$ is easily derived
\begin{equation} [{\cal R}_{i},{\cal R}_{j}]+[{\cal R}_{i},{\cal
L}_{j}]=2\epsilon_{ij}^{l}{\cal R}_{l} . \label{adjrepR} \end{equation}

We now have derived the basic representations (adjoint and fundamental) of
particle physics for alternative algebras.

\section{An octonionic gauge theory}

Having developed both fundamental and adjoint representations we can now
proceed to write a ``genuine'' octonionic gauge theory. However there are new
features peculiar to the representation theory of alternative algebras which
dictate changes to the usual Lie group gauge theories. We now have 7 left and
right representation matrices~\cite{tau} denoted $L_{i}$ and $R_{j}$ where
$i=1,\ldots ,7$. Denote the identity matrix ${\bf 1}=L_{0}=R_{0}$ so that
$(L\{R\}_{\alpha})=(L\{R\})=({\bf 1},L\{R\}_{i})$. The extended structure
constants $c_{\alpha\beta\gamma}$ can be read directly from the octonion
multiplication table including the identity.

 From the defining relations~\ref{repreln} we have
\begin{eqnarray}
L_{\alpha}L_{\beta} + [L_{\alpha},R_{\beta}] & = &
c_{\alpha\beta\gamma}L_{\gamma} \label{Left}\\
R_{\alpha}R_{\beta} + [R_{\alpha},L_{\beta}] & = &
c_{\beta\alpha\gamma}R_{\gamma} \label{Right}\\
\! [L_{\alpha},R_{\beta}] & = &  [R_{\alpha},L_{\beta}]
\end{eqnarray}

These relations mimic those for a Lie algebra save for the additional
non-linear commutator terms linking the left and right sides of the
representation.  If we consider just left transformations on a matter field
$\psi$ of the usual type \begin{equation}\psi \rightarrow \psi ' =
e^{i\vec{\alpha}.\vec{L}}\psi\end{equation} then we incur a right
transformation via the linking commutator terms. This failure of either side of
the representation algebra to close is an essential feature of any
non-associative gauge theory. In essence the undesirable non-associativity of
the algebra has been reformatted as a closure problem in our Lie Algebra
structure.

A first solution to the closure problem is to add such generators to the left
(or right) matrix generators necessary to close the algebra. One is then
dealing with an ordinary Lie Algebra with a subalgebra of particular interest
representing the octonion symmetry. Although this approach is not completely
valueless, since one would expect it to provide some of the features of an
octonionic symmetry, we feel that it does not fit the criteria of a ``genuine''
octonionic gauge theory. Rather we would prefer to invoke new physics to handle
the inherent closure problem for the strong association between the octonions
and Lie groups is well known~\cite{GCJ}.

Consider $SU(2)$ internal symmetry space. The tangent space at each point is a
three dimensional complex vector space spanned by the Pauli matrices with
multiplication law
\begin{equation}[\sigma_{i},\sigma_{j}]=2\epsilon_{ijk}\sigma_{k} \ \ \mbox{
where } a,b=1,2.\end{equation}
Suppose we were now to isolate two of these complex directions so as to operate
only with the first two Pauli matrices $\sigma_{1}$ and $\sigma_{2}$ (this
choice is arbitrary). The algebra of the generators~\cite{fn4} of this two
dimensional complex vector space is obviously not closed since
\begin{equation}[\sigma_{a},\sigma_{b}]=2\epsilon_{ab}M \ \ \mbox{ where }
a,b=1,2\end{equation}
and $\epsilon_{ab}$ is the two dimensional alternating symbol. The ``unseen''
generator $M$ corresponds to the third Pauli matrix $\sigma_{3}$. There is,
however no impediment to the usual $SU(2)$ calculations. Rather, choosing to
operate with only two of the three available complex dimensions will produce
some peculiar but artificial effects.

The story for the octonions is more subtle since by virtue of our associative
matrix representation we must encounter a non-closed algebra. Yet the
generators of our fundamental representation are also generators of the $SO(8)$
symmetry group. Hence we handle the failure of closure for our representation
of the octonion algebra by carrying out all calculations with generators of the
$SO(8)$ symmetry group but isolate those generators corresponding to the
octonion algebra. So in a sense an octonionic symmetry implies an overall
$SO(8)$ symmetry of which only specific channels are ever observed. Conversely
one can take the viewpoint that we have found an embedding of the octonion
symmetry in $SO(8)$~\cite{fn5}. We now present the skeletal structure of an
octonionic Yang-Mills theory.

Begin with a matter field
\begin{equation}\Psi = (\psi^{\alpha})= \left( \begin{array}{c} \psi^{0} \\
\psi^{1} \\ \psi^{2} \\ \psi^{3} \end{array}   \right) \end{equation}
where the entries $\psi^{\alpha}$ are complex spinors. Firstly notice that if
the $\psi^{\alpha}$ transform separately under the fundamental representation
of $SU(2)$ \begin{equation}\psi^{\alpha}\rightarrow
e^{\frac{i}{2}\vec{\theta}(x).\vec{\sigma}}\psi^{\alpha}\end{equation} so that
the Lagrangian of the theory is invariant ${\cal L}(\Psi ) \rightarrow
e^{-\frac{i}{2}\vec{\theta}(x).\vec{\sigma}}{\cal L}(\Psi
)e^{\frac{i}{2}\vec{\theta}(x).\vec{\sigma}}$ then we trivially have an
octonionic symmetry since the left and right representation matrices are
composed of Pauli matrices. Let us assume that this is not the case but rather
consider a full octonionic symmetry.

As is usual for Lie algebras let us consider only left global symmetry
transformations of the type
\begin{equation}\Psi(x)\rightarrow \Psi '(x)=e^{\frac{i}{2}\vec{\alpha}(x)
.\vec{\lambda}}\Psi(x) \ \mbox{ with }
\vec{\alpha}.\vec{\lambda}=\sum_{i=1}^{7}\alpha_{i}\lambda_{i}\end{equation}
where the $\lambda$ and $\rho$ matrices $\lambda_{i}=iL_{i}$ and $\rho=iR_{i}$
are Hermitean. However in doing so we now encounter a non-closed algebra. The
commutation and anti-commutation relations for the $\lambda_{i}$ are
\begin{eqnarray}
[\frac{\lambda_{i}}{2},\frac{\lambda_{j}}{2}] \
&=&\epsilon_{ijk}\frac{\lambda_{k}}{2}+ [\rho_{j},\frac{\lambda_{i}}{2}]
\label{comm}\\
\{ \frac{\lambda_{i}}{2},\frac{\lambda_{j}}{2} \} &=& \delta_{ij} \bf{1/2}.
\end{eqnarray}
The non-linear left-right commutator term in the commutation relations reflects
the non-associative triple product structure of the octonions and will give new
terms in the curvature tensor as yet unseen for Lie algebras.

Now construct a covariant derivative ${\cal D}_{\mu}$ ($\mu$ is a Lorentz
spacetime index) so that the variation of ${\cal D}_{\mu}\Psi$ has the same
transformation as $\Psi$ itself
\begin{equation}{\cal D}_{\mu}\Psi\rightarrow ({\cal D}_{\mu}\Psi )' =
e^{\frac{i}{2}\vec{\alpha}(x) .\vec{\lambda}}{\cal D}_{\mu}\Psi.\end{equation}
We therefore introduce a gauge field ${\bf A}_{\mu}$ in the usual fashion so
that
\begin{equation}{\cal D}_{\mu}\Psi
=(\partial_{\mu}-ig\frac{\vec{\lambda}.\vec{{\bf
A}_{\mu}}}{2})\Psi.\end{equation}
The transformation of the gauge field is again familiar
\begin{equation}\frac{\vec{\lambda}.\vec{{\bf A}_{\mu}}}{2}\rightarrow
(\frac{\vec{\lambda}.\vec{{\bf A}_{\mu}}}{2})'=e^{\frac{i}{2}\vec{\alpha}(x)
.\vec{\lambda}}(\frac{\vec{\lambda}.\vec{{\bf
A}_{\mu}}}{2})e^{-\frac{i}{2}\vec{\alpha}(x)
.\vec{\lambda}}-\frac{i}{g}(\partial_{\mu}e^{\frac{i}{2}\vec{\alpha}(x).\vec{\lambda}})e^{-\frac{i}{2}\vec{\alpha}(x).\vec{\lambda}}\end{equation}
To first order using the commutation relation~\ref{comm} we have
\begin{equation}(\frac{\vec{\lambda}.\vec{{\bf A}_{\mu}}}{2})'=
\frac{\vec{\lambda}.\vec{{\bf
A}_{\mu}}}{2}-i\alpha^{j}A^{k}_{\mu}(\epsilon_{ijk}\lambda_{i}+[\rho_{j},\frac{\lambda_{k}}{2}]) -\frac{\vec{\lambda}}{2g}.\partial_{\mu}\vec{\alpha}\label{gauge}\end{equation}
The second term is just the transformation law for a septet under the left side
of the adjoint representation described in
 section~\ref{adj}. Non-associativity invokes the appearance of the rank two
tensor transformation term
$\frac{1}{2}\alpha^{j}A^{k}_{\mu}[\rho_{j},\lambda_{i}]$. We may divide the
transformation~\ref{gauge} into closed and non-closed algebraic parts by
denoting
 ${A^{i}}_{\mu}'=A^{i}_{\mu}-i\epsilon_{ijk}\alpha^{j}A^{k}_{\mu}
-\frac{1}{g}\partial_{\mu}\alpha^{i}$ (where of course
$\vec{\lambda}.\vec{A}_{\mu}=\lambda_{i}A_{\mu}^{i}$) so that
\begin{equation}(\frac{\vec{\lambda}.\vec{{\bf
A}_{\mu}}}{2})'=\frac{1}{2}{A_{\mu}^{i}}'L_{i}
+\alpha^{j}A^{k}_{\mu}[\rho_{j},\frac{\lambda_{k}}{2}]\end{equation} where the
second term carries the non-associativity of the algebra failing to close back
onto the algebra of the $\lambda$ matrices. A non-linear term in the
anti-symmetric curvature tensor emerges similarly. We find
\begin{equation}({\cal D}_{\mu}{\cal D}_{\nu}-{\cal D}_{\nu}{\cal
D}_{\mu})=\frac{ig}{2}(\lambda_{i}F^{i}_{\mu\nu}-A_{\mu}^{j}A_{\nu}^{k}[\rho_{j},\lambda_{k}])\equiv {\cal F}_{\mu\nu}\end{equation}
where the associative term is as usual
\begin{equation}F^{i}_{\mu\nu}=\partial_{\mu}A^{i}_{\nu}-\partial_{\nu}A^{i}_{\mu}-ig\epsilon_{ijk}A^{j}_{\mu}A^{k}_{\nu}.\end{equation}

Finally let us now consider the form of a source-less Yang-Mills Lagrangian
\begin{equation}{\cal L}=tr({{\cal F}_{\mu\nu}{\cal
F}^{\mu\nu}}).\end{equation}
We need the following trace relations
\begin{eqnarray}
tr(\lambda_{i}\lambda_{j})&=&8\delta_{ij} \\
tr(\lambda_{i}[\rho_{j},\lambda_{k}])&=& 8\epsilon_{ijk}\\
tr([\lambda_{i},\rho_{j}][\lambda_{k},\rho_{l}])&=&8[\epsilon_{ijkl}+2(\delta_{ik}\delta_{jl}-\delta_{il}\delta_{jk})].
\end{eqnarray}

Therefore we have
\begin{equation}{\cal
L}=-g^{2}[F^{i}_{\mu\nu}F^{i\mu\nu}-2\epsilon_{ijk}A^{i}_{\mu}A^{j}_{\nu}F^{k\mu\nu}+(\epsilon_{ijkl}+2(\delta_{ik}\delta_{jl}-\delta_{il}\delta_{jk}))A^{i}_{\mu}A^{j}_{\nu}A^{k\mu}A^{l\nu}].\end{equation}
The first term is the usual associative kinetic term but now non-associativity
leads to a bilinear coupling between the gauge fields and curvature tensor and
a term quadrilinear in the gauge fields. These terms provide new interactions,
the phenomenology of which is currently being pursued.
\section{Summary and Conclusion}
Although no faithful matrix representation of the octonions can exist since
matrices associate whereas the octonions do not, like any algebra their
structure constants and division tables can be used to construct matrix
representatives for the octonion algebra. Representation theory for alternative
algebras has existed in mathematics since early this century. However we have
shown that it provides a fundamental and adjoint representation for the
octonion algebra. Furthermore an associative representation theory resolves the
unitarity problem usually thwarting octonionic quantum mechanics.
Non-associativity of course could not be made to disappear so easily but rather
rears its head as a failure of the symmetry algebra to close giving rise to new
terms in the transformation laws for matter and gauge fields and in the kinetic
term of a gauge invariant Lagrangian. The phenomenology of such terms promises
to be of interest especially since the octonions themselves contain so much
structure. Therefore the possibi
lity of an octonionic quark theory still, flickeringly, beckons.

\section{Acknowledgements}
The authors would like to thank J. Choi, J. Daicic, A. Davies, C. Dettmann and
B. Hanlon for stimulating discussions. One of us (A.K.W.) would like to
acknowledge the support of the Australian Postgraduate Research Program and the
Dixon Scholarship. Finally the feline interactions of Aldous Huxley and Olga
were invaluable.

\newpage
\section{Appendix}\label{appendix}
\subsection{Octonionic division tables}
\begin{center} \begin{tabular}{c|cccccccc}Left & & & & & & & & \\ $\div$ &
$e_{0}$ &$e_{1}$  & $e_{2}$ &$e_{3}$  &$e_{4}$  &$e_{5}$  &$e_{6}$ & $e_{7}$
\\
	\hline			$e_{0}$ &	$e_{0}$ &$-e_{1}$  & $-e_{2}$ &$-e_{3}$  &$-e_{4}$
&$-e_{5}$  &$-e_{6}$  &$-e_{7}$   \\
	$e_{1}$ &$e_{1}$  & $e_{0}$ &$-e_{3}$  &$e_{2}$  &$-e_{5}$  &$e_{4}$  &$e_{7}$
 &$-e_{6}$ \\
	$e_{2}$ &$e_{2}$  & $e_{3}$ &$e_{0}$  &$-e_{1}$  &$-e_{6}$  &$-e_{7}$
&$e_{4}$  &$e_{5}$ \\
	$e_{3}$ &$e_{3}$  & $-e_{2}$ &$e_{1}$  &$e_{0}$  &$-e_{7}$  &$e_{6}$
&$-e_{5}$  &$e_{4}$ \\
	$e_{4}$ &$e_{4}$  & $e_{5}$ &$e_{6}$  &$e_{7}$  &$e_{0}$  &$-e_{1}$  &$-e_{2}$
 &$-e_{3}$ \\
	$e_{5}$ &$e_{5}$  & $-e_{4}$ &$e_{7}$  &$-e_{6}$  &$e_{1}$  &$e_{0}$  &$e_{3}$
 &$-e_{2}$ \\
	$e_{6}$ &$e_{6}$  & $-e_{7}$ &$-e_{4}$  &$e_{5}$  &$e_{2}$  &$-e_{3}$
&$e_{0}$  &$e_{1}$ \\
	$e_{7}$ &$e_{7}$  & $e_{6}$ &$-e_{5}$  &$-e_{4}$  &$e_{3}$  &$e_{2}$
&$-e_{1}$  &$e_{0}$ \\
\end{tabular} \end{center}
\begin{center} \begin{tabular}{c|cccccccc}Right & & & & & & & & \\ $\div$ &
$e_{0}$ &$e_{1}$  & $e_{2}$ &$e_{3}$  &$e_{4}$  &$e_{5}$  &$e_{6}$ & $e_{7}$
\\
	\hline			$e_{0}$ &	$e_{0}$ &$-e_{1}$  &$-e_{2}$ &$-e_{3}$  &$-e_{4}$
&$-e_{5}$  &$-e_{6}$  &$-e_{7}$   \\
	$e_{1}$ &$e_{1}$  & $e_{0}$ &$ e_{3}$  &$-e_{2}$  &$ e_{5}$  &$-e_{4}$
&$-e_{7}$  &$ e_{6}$ \\
	$e_{2}$ &$e_{2}$  & $-e_{3}$ &$e_{0}$  &$ e_{1}$  &$ e_{6}$  &$ e_{7}$
&$-e_{4}$  &$-e_{5}$ \\
	$e_{3}$ &$e_{3}$  & $ e_{2}$ &$-e_{1}$  &$e_{0}$  &$ e_{7}$  &$-e_{6}$  &$
e_{5}$  &$-e_{4}$ \\
	$e_{4}$ &$e_{4}$  & $-e_{5}$ &$-e_{6}$  &$-e_{7}$  &$e_{0}$  &$ e_{1}$  &$
e_{2}$  &$ e_{3}$ \\
	$e_{5}$ &$e_{5}$  & $ e_{4}$ &$-e_{7}$  &$ e_{6}$  &$-e_{1}$  &$e_{0}$
&$-e_{3}$  &$ e_{2}$ \\
	$e_{6}$ &$e_{6}$  & $ e_{7}$ &$ e_{4}$  &$-e_{5}$  &$-e_{2}$  &$ e_{3}$
&$e_{0}$  &$-e_{1}$ \\
	$e_{7}$ &$e_{7}$  & $-e_{6}$ &$ e_{5}$  &$ e_{4}$  &$-e_{3}$  &$-e_{2}$  &$
e_{1}$  &$e_{0}$ \\
\end{tabular} \end{center}
\newpage

\begin{references}
\bibitem[5]{fn1}Notice that the concept of a soft Lie algebra is really just an
extension of the theory of deformations of Lie algebras.
\bibitem[7]{pauli}Notice that if we had chosen to use the right division table
we would have obtained the usual correspondance $i \leftrightarrow
-i\sigma_{1}$, $j \leftrightarrow -i\sigma_{1}$, $k \leftrightarrow
-i\sigma_{1}$. This arbitrariness in sign of course has no effect on actual
physics.
\bibitem[8]{fn2}Although this could be a useful property if one wants to
consider systems where the observer as well as the system is quantum.
\bibitem[10]{fn3}There is of course a temptation to call the last of these the
``Adams relation''.
\bibitem[11]{tau}Notice that there are 7 other natural rank one matrix
generators in the theory $\tau_{i}\equiv \epsilon_{ijk}[L_{j},R_{k}]$. The
triple product structure of the octonions necessitates the inclusion of the
rank two matrix generators $[L_{i},R_{j}]$.
\bibitem[12]{fn4}This is a concept often utilised in spacetime so there is no
reason why we should not extend it to symmetry space.
\bibitem[13]{fn5}Also note that this scheme is not to mistaken for symmetry
breaking. Instead it is really the only sensible avenue remaining when one
wants to work with a manifestly non-closed algebra. (That non-associativity is
closely tied to non-closure can be seen to be natural if one remembers that
non-associativity is equivalent to the failure of transitivity in an
equivalence relation).
\end{references}
\end{document}